# Scanning SQUID-on-tip microscope in a top-loading cryogen-free dilution refrigerator

Haibiao Zhou[1#], Nadav Auerbach[1#], Indranil Roy[1], Matan Bocarsly[1], Martin E. Huber[2], Barun Barick[1], Arnab Pariari[1], Markus Hücker[1], Zhi Shiuh Lim[3], A. Ariando[3], Alexey I. Berdyugin[4,5], Na Xin[4,5], Michael Rappaport[6], Yuri Myasoedov[1], and Eli Zeldov[1*]

[1]Department of Condensed Matter Physics, Weizmann Institute of Science, Rehovot 7610001, Israel
[2]Departments of Physics and Electrical Engineering, University of Colorado Denver, Denver, CO, USA
[3]Department of Physics, National University of Singapore, Singapore
[4]School of Physics and Astronomy, University of Manchester, Manchester M13 9PL, UK
[5]National Graphene Institute, University of Manchester, Manchester M13 9PL, UK
[6]Department of Physics Core Facilities, Weizmann Institute of Science, Rehovot 7610001, Israel
# These authors contributed equally
* eli.zeldov@weizmann.ac.il

**Abstract**

The scanning superconducting quantum interference device (SQUID) fabricated on the tip of a sharp quartz pipette (SQUID-on-tip) has emerged as a versatile tool for nanoscale imaging of magnetic, thermal, and transport properties of microscopic devices of quantum materials. We present the design and performance of a scanning SQUID-on-tip microscope in a top-loading probe of a cryogen-free dilution refrigerator. The microscope is enclosed in a custom-made vacuum-tight cell mounted at the bottom of the probe and is suspended by springs to suppress vibrations caused by the pulse tube cryocooler. Two capillaries allow in-situ control of helium exchange gas pressure in the cell that is required for thermal imaging. A nanoscale heater is used to create local temperature gradients in the sample, which enables quantitative characterization of the relative vibrations between the tip and the sample. The spectrum of the vibrations shows distinct resonant peaks with maximal power density of about 27 nm/Hz$^{1/2}$ in the in-plane direction. The performance of the SQUID-on-tip microscope is demonstrated by magnetic imaging of the MnBi$_2$Te$_4$ magnetic topological insulator, magnetization and current distribution imaging in a SrRuO$_3$ ferromagnetic oxide thin film, and by thermal imaging of dissipation in graphene.

**Introduction**

Nanoscale imaging with scanning probe microscopes (SPM) can provide indispensable information for the understanding of quantum materials and devices. The scanning SQUID microscope has been demonstrated to be an ultra-sensitive magnetic imaging tool and an important technique in the study of superconductivity, magnetism, and topological materials[1–7]. However, due to the large size of the SQUID pickup loops fabricated using conventional lithographic methods, the spatial resolution is usually limited to about 1 µm, which is too large for many applications, especially in devices with micrometer sizes. Many efforts have been made to miniaturize the SQUID pickup loops[8,9]. This limitation is largely alleviated by the recently developed SQUID-on-tip (SOT) technique[10,11], which utilizes a three-step self-aligned deposition of superconducting film on a pulled quartz tube to produce a nanoscale SQUID[12]. The effective diameter of the SQUID, which resides at the apex of the quartz pipette, can be as small as 40 nm[13]. Another advantage of this geometry is that it allows a tip-sample distance as small as 10 nm, limited by system vibrations. The magnetic sensitivity of the SOT can reach a single electron spin[12], and in addition, by exploiting the temperature dependence of the critical current, thermal imaging of the dissipation processes at the nanoscale can be attained[14] with unprecedented thermal sensitivity of better than 1 µK/Hz$^{1/2}$. This technique has been recently applied to various materials and has revealed multiple phenomena, including vortex dynamics in superconductors[15–18], atomic-scale dissipation due to resonant scattering in graphene[14,19,20], imaging equilibrium currents in the quantum Hall state[21,22], nanoscale magnetism in oxides[23], magnetic topological insulators[24] and van der Waals ferromagnets[25], and orbital magnetism in moiré superlattices[26–28]. The technique has also been extended to the multiterminal SOT with a tunable interference pattern[29], as well as the SQUID-on-lever providing simultaneous topographic imaging[30].

Due to the high cost of liquid helium (He) and interruptions in its supply, cryogen-free cryostats and dilution refrigerators (DR) are becoming increasingly popular in low-temperature SPM systems[31–37] including scanning tunneling microscopes, which have the highest requirement for vibration isolation[38–43]. Other advantages include continuous operation without interruption for cryogen transfers. However, these benefits come at the price of introducing mechanical and acoustic vibrations to the system. Most of the dry DRs used for SPM systems have a larger available sample space as compared to wet DRs, which readily allows the use of springs for vibration isolation[31,35,38]. In such setups, however, the DR system has to be heated to room temperature for exchange of sensor or sample, which is time consuming. The alternative top- or bottom-loading design has the advantage of a faster turn-around time, but puts stringent limitations on the microscope design and especially for vibration isolation due to the limited space. Hence, SPMs in a top/bottom-loading dry DR are rare[37]. Here, we report the design and performance of a scanning SOT microscope in a top-loading dry DR. A unique feature of the system is a vacuum-tight cell that encloses the microscope, which allows in-situ control of the pressure of the He exchange gas that is required for thermal imaging, including immersion of the microscope in liquid He. We show that by using inertia blocks for cryostat support and spring suspension of the microscope, low vibrations suitable for scanning can be achieved. High quality magnetic and thermal imaging are presented as demonstration of the system performance.

**Microscope design**

Our dry DR system is model CF1200 from Leiden Cryogenics equipped with a vector magnet of 5 T along the $z$ axis and a rotatable field vector of 1 T. The cryocooler is a Cryomech PT415 pulse tube with cooling power of 1.5 W at 4.2 K. The cooler generates gas pulses at 1.4 Hz, which is the main source of vibrations. The DR has a top-loading design, with a Ø50 mm cold insertable probe with a vacuum load lock. The probe provides access to all the cold plates, which is needed for housing the cold preamplifier and facilitates further modification. It has four sets of clamping arms for thermal anchoring to the four cooling stages of the DR. The first challenge is the small diameter of the probe, which limits the space for the microscope and the vibration isolation. The second challenge is that the probe resides within the inner vacuum chamber (IVC) of the cryostat and is hence under vacuum conditions. This is not a limitation for magnetic imaging; however, thermal imaging requires finite He exchange gas pressure for thermal coupling of the SOT to the sample[14]. Hence, the microscope needs to be enclosed within a separate vacuum-tight chamber in which a stable and controllable pressure can be maintained. This poses an additional space limitation and requires leak-tight cryogenic electrical feedthroughs and gas capillaries.

The cryostat is mounted on a triangular shaped concrete inertia block with a weight of 2 tons (Fig. 1a), which resides over a concrete pit. The inertia block can be suspended by three pneumatic vibration isolators. Since the resonance frequency of the isolators of about 1 Hz is close to the 1.4 Hz frequency of the pulse tube, the suspension causes an overall vibration of the inertia block. Hence all the presented measurements were conducted without block suspension. The linear motor of the cryocooler is firmly fixed to a separate concrete stage, while the remote compressor is connected to the linear motor through 15 m flexible stainless-steel hoses. The gas handling system (GHS) with the vacuum pumps is placed on a suspended concrete block in a neighboring room and the mixture pumping line is anchored to a concrete wall.

Figure 1b shows the schematics of the bottom part of the probe and the main custom modifications. The cell enclosing the microscope is firmly attached to the mixing chamber plate through a heat sink made of brass. The indium-sealed top flange of the cell acts as a feedthrough for electrical ports, two gas capillaries, and three Cu cold fingers. This custom-designed vacuum-tight feedthrough (Winchester Interconnect) contains a 51-pin micro-D connector, 9 coaxial SMP connectors and one triax connector (Fig. 1e). A photo of the bottom end of the probe and the scan head with the cell removed is shown in Fig. 1c. To flush the cell and control the He pressure, we use two stainless steel capillaries of 1 mm inner diameter and 0.3 mm wall thickness that run from the top of the probe to the 3 K plate, while from 3 K plate to the cell we use thinner capillaries with 0.5 mm inner diameter and 0.15 mm wall thickness to reduce heat leaks between the different plates. They are well thermally anchored at all stages by soldering a 20 cm section of the capillary that is wounded around bobbins made of oxygen-free high thermal conductivity copper (Fig. 1c). Two vacuum gauges are connected to the room-temperature end of the capillaries to monitor the pressure inside the cell. A flow meter is used to monitor and control the flow rate and total volume of the He gas introduced into the cell. The heat sink has three long copper cold fingers that penetrate into the cell through the three tubes of the feedthrough to help cool down the cell (Fig. 1d). Both the cold fingers and the capillaries are welded to the feedthrough using silver brazing.

The scanning microscope (Fig. 2c) consists of a tip puck with coarse steppers (Attocube ANPx101 and ANPz101) and a sample puck with scanner (Attocube ANSxyz100), connected by a sleeve. The travel range of the steppers is 5 mm in all directions and a scan range is about 30 x 30 μm². The two pucks and the sleeve are made of Grade 2 Titanium. The microscope is hung in the cell by three springs (BeCu, spring constant = 15 N/m, free length = 50 mm) for vibration isolation, which help attenuate the vibration with frequencies above 10 Hz. The microscope is thermally anchored to the cold fingers of the heat sink by copper braids, which also help to dampen the oscillations to prevent resonance amplification. We use the Nanonis SPM controller from SPECS GmbH with ANC250 (Attocube) high voltage amplifier to control the scanning.

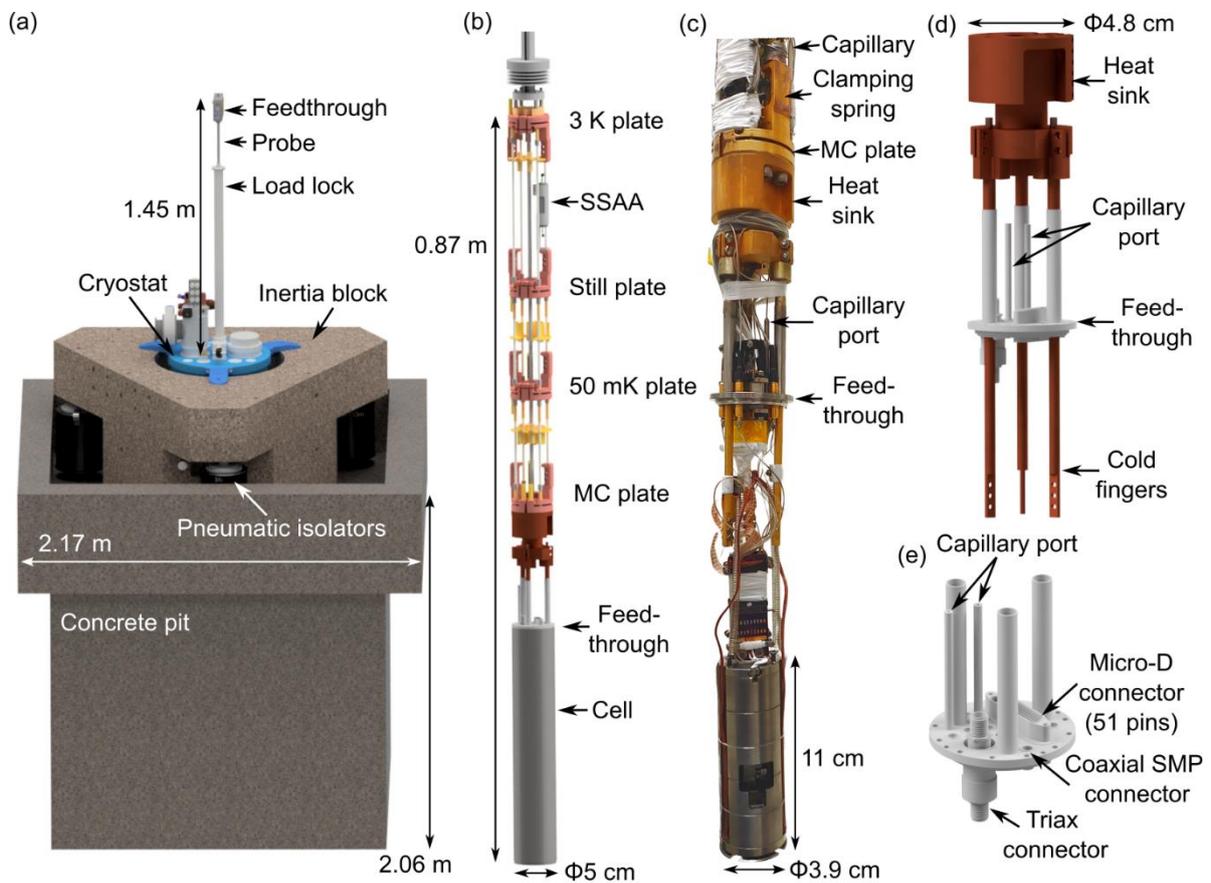

**Fig. 1**. (a) Drawing of the cryogen-free cryostat, the top-loading probe, and the supporting concrete structure. (b) 3D model of the lower part of the probe and the cell. (c) A photograph showing the bottom part of the probe with the microscope. (d) 3D model of heat sink and feedthrough. (e) Model of the vacuum-tight feedthrough.

The fabrication process of the SOT, either by thermal evaporation or magnetron sputtering, has been described elsewhere[10,11,44]. A typical SEM image of a Pb tip is shown in Fig. 2a. SOTs with different diameters from 50 to 250 nm were fabricated. The SOT readout circuit is based on the SQUID series array amplifier (SSAA)[45], which amplifies the current through the SOT with a gain of about 13.5, then with a current to voltage gain of 1 kΩ. The interference pattern of the same Pb SOT is shown in Fig. 2d. The SSAA, placed in a magnetic field shielding enclosure, is firmly fixed and thermally anchored to the still plate on the probe (Fig. 1b).

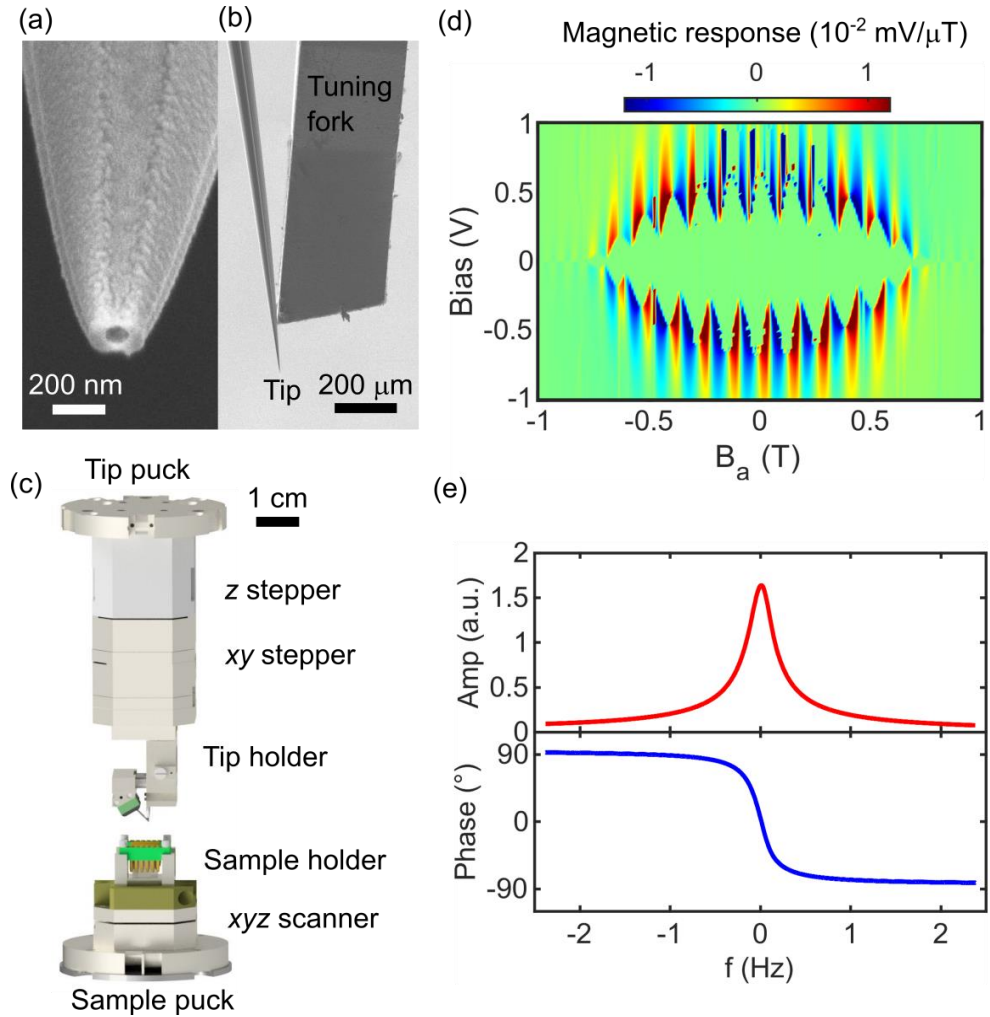

Fig. 2. (a) SEM image of a typical 140 nm diameter Pb SOT. (b) Side view of the SOT pipette attached to one tine of a quartz tuning fork for force sensing. The tuning fork vibrates in the direction perpendicular to the plane of the image. (c) The scanning SOT microscope consisting of the tip puck with steppers and sample puck with scanner. The sleeve connecting the tip puck and sample puck is not shown. (d) The magnetic response $dV_{out}/dB_a$ of the same Pb tip vs. the applied magnetic field $B_a$ and the SOT circuit bias voltage (V), measured at 4.6 K. $V_{out}$ is the voltage output of the SOT amplification circuit. (e) Measured resonance response of the tuning fork showing the amplitude (top) and the phase (bottom) vs. the excitation frequency. $f$ is the frequency shift from the resonance frequency of 32.8 kHz.

A quartz tuning fork (TF) with a resonant frequency of about 32.8 kHz, is used for tip-sample distance control. The SOT is pressed against one tine of the TF with a protrusion of about 100 μm (Fig. 2b). A SPECS OC4 oscillation controller is used to drive the TF electrically and detect the phase change. A room temperature transimpedance amplifier with a gain of 10 MΩ, adapted from ref. [46], is used for the measurement of the TF signal. The tuning fork vibration amplitude is usually several tens of nanometers with a *Q* factor of up to 120,000 (Fig. 2e). This also enables controllable vibration of the SOT parallel to the sample surface for direct measurement of the gradient of the magnetic or thermal signal along the vibration axis with high signal-to-noise ratio.

**Vibration measurements**

The pulse tube cryocooler inevitably introduces relative vibrations between the tip and the sample, which can be quantitatively characterized using a technique recently developed by Schiessl *et al.*[47,33,36]. In this method, a highly localized *dc* magnetic field, created for example by a superconducting vortex, is imaged while a noise spectrum is recorded at each pixel. By calculating the 3D gradients of the magnetic field, the amplitude and the direction of the vibrations can be derived from the position-dependent noise signal. Rather than using a localized magnetic signal, we have generated a localized heat signal by driving a current of 20 μA through a 200 nm long and narrow constriction made of thin gold film. Figure 3a shows the image of the local temperature variation $T(x,y)$ acquired by a Mo$_{66}$Re$_{34}$ ⌀174 nm SOT scanned at a height $h = $ 100 nm above the sample surface at 1.5 K in a He pressure of 5 mbar. By acquiring another scan at a slightly different height, $h + \Delta h$, the temperature derivative along the $z$ axis, $\partial T(x,y,z)/\partial z \approx \Delta T(x,y,z)/\Delta h$, was attained, where $\Delta T(x,y,z)$ is the numerical subtraction of the two scans (Fig. 3b). In addition, the spatial derivatives along the $x$ and $y$ axes, $\partial T/\partial x$ and $\partial T/\partial y$, are calculated numerically as shown in Fig. 3c-d.

During the scanning, the noise spectrum of the temperature signal, $\mathcal{T}^{1/2}(f,x,y,h)$, was recorded at each pixel in a frequency window of up to $f = $ 1.5 kHz using a spectrum analyzer (SR1, Stanford Research Systems). An example of the noise spectrum at one location is shown in Fig. 3e. Apart from the 50 Hz noise and its harmonics, several strong peaks are present and labelled by the arrows. Left column of Fig. 3g shows the spatial map of $\mathcal{T}_f^{1/2}(x,y)$ at several frequencies corresponding to the peaks in the spectrum. In the presence of mechanical vibrations, the temperature gradients translate the vibrations into SOT noise given by

$$|\mathcal{T}_f^{1/2}(f,x,y,h)| = \left| \frac{\partial T(x,y,h)}{\partial x}\delta x(f) + \frac{\partial T(x,y,h)}{\partial y}\delta y(f) + \frac{\partial T(x,y,z)}{\partial z}\delta z(f) \right|, \quad (1)$$

where the three gradient terms are given by the gradient maps presented in Fig. 3b-d. The vibration amplitudes at frequency *f* along the three axes, $\delta x(f)$, $\delta y(f)$ and $\delta z(f)$, can then be derived by performing a three-parameter least-squares fit. The middle column of Fig. 3g shows the fitting and the derived vibration amplitudes $\delta z$ and $\delta r = (\delta x^2 + \delta y^2)^{1/2}$ for the different frequencies, along with their difference from the measured noise (right column). This procedure has been performed for each frequency of the discrete spectrum resulting in the vibration amplitude spectra of $\delta z$ and $\delta r$ presented in Fig. 3f.

The results show that the vibrations have prominent amplitudes at several frequencies marked by the arrows in Fig. 3e. The in-plane component of the vibrations $\delta r$ (red in Fig. 3f) is found to be usually much larger than the out-of-plane component $\delta z$ (blue), similar to the reports of other dry cryostat systems[47,33,34,36]. The in-plane vibrations are found to be mainly along two directions as marked by the white arrows in Fig. 3g. The low frequency vibrational peaks at 11.8 and 21.6 Hz are likely the harmonics of the 1.4 Hz He gas pulses, while the 179.8 Hz and higher peaks apparently arise from the resonant frequencies of the dewar and the piezo scanners and steppers. The largest vibration amplitude of $\delta r \cong$ 27 nm/Hz$^{1/2}$ is found at $f =$ 179.8 Hz. The above analysis provides the means for distinguishing between electrical and mechanical noise sources, thus facilitating system troubleshooting and optimization. The peaks marked by the arrows in Fig. 3e arise from mechanical vibrations as reflected by Fig. 3f, while the rest of the peaks arise from electrical sources as exemplified by the 50 Hz, 100 Hz and 857 Hz data in Fig. 3h. The vibrations restricted our practical minimal scanning height to $h =$ 40 nm.

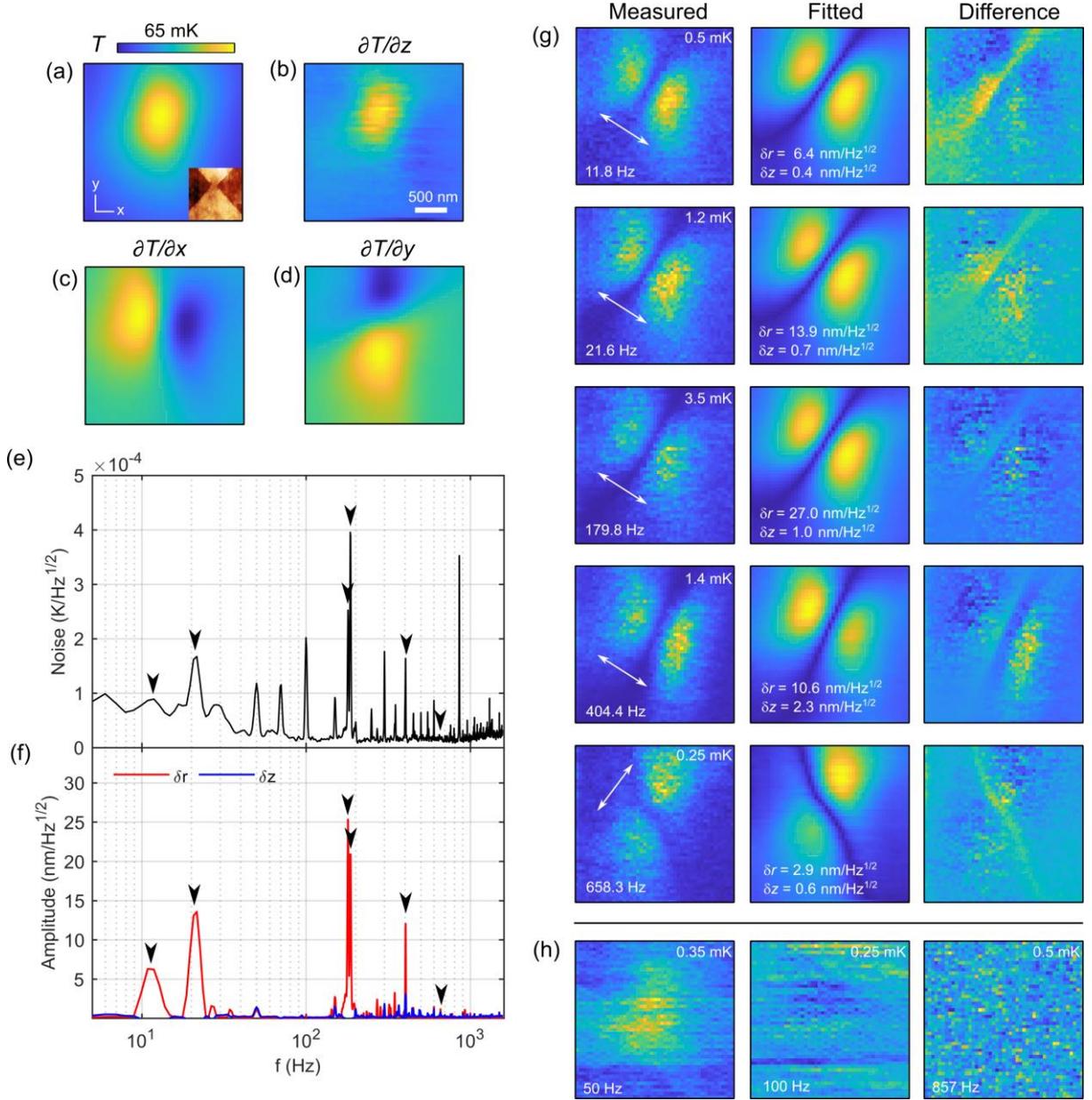

**Fig. 3.** (a) Image of the local temperature variation $T(x,y)$ acquired with a Mo$_{66}$Re$_{34}$ SOT at 1.5 K at a scan height $h = 100$ nm. The inset shows the atomic force microscopic image of the constriction with the same scan area. (b) $\partial T(x,y,z)/\partial z$ calculated by subtracting two thermal images acquired at heights of $h = 100$ and 150 nm. (c) $\partial T(x,y)/\partial x$ calculated by taking a numerical derivative of (a) along the $x$ axis. (d) Calculated $\partial T(x,y)/\partial y$. (e) Noise power density measured by the SOT at one of the tip positions (at the bottom left corner of the image) presented in units of temperature spectral density K/Hz$^{1/2}$. The arrows mark the dominant noise peaks arising from mechanical vibration resonances. (f) Spectrum of the mechanical SOT vibrations relative to the sample derived from the analysis of the noise images as presented in (g). The in-plane vibration amplitude $\delta r$ (red) is much larger than the out-of-plane vibration $\delta z$ (blue). The arrows mark the peaks arising from mechanical vibrations while the rest of the peaks in (e) arise from electrical noise. (g) Images of the position-dependent noise power density at different indicated frequencies (left column); fit to the noise distribution $|\mathcal{T}_f^{1/2}(x,y)|$ using Eq. (1) with indicated best fit parameters $\delta r$ and $\delta z$ (center

column); difference between the measured noise and the fitted data (right column). The white arrows in the left column show the derived direction of the in-plane vibrations. (h) The 50 Hz, 100 Hz and 587 Hz data demonstrating noise arising from electrical source rather than mechanical vibrations. The colormap range for the measured data in (g) and (h) is indicated at the top right corner of the images. Size of all images: 2.5 x 2.5 μm².

**Demonstration of thermal and magnetic imaging**

The base temperature of the DR probe at the mixing chamber plate is about 100 mK, while the sample temperature is about 150 mK measured with a ruthenium oxide thermometer mounted on the sample holder. When imaging large areas with scanning speed of 0.5 μm/s, the dissipation in the piezo scanners causes the sample temperature to increase to about 250 mK. We believe this temperature increase can be reduced by improving the thermal anchoring of the scanners.

For magnetic imaging we have employed either Pb ($T_C \cong$ 7.2 K) or In ($T_C \cong$ 3.5 K) SOTs. The magnetic sensitivity of the SOTs can be as high as 10 nT/Hz$^{1/2}$ in applied magnetic fields $(B_a)$ up to 1.5 T for Pb SOTs at base temperature. The magnetic imaging can be performed either in vacuum or in He exchange gas from base temperature up to $T_C$. For thermal imaging, however, the thermal conduction of the He gas below 1.2 K becomes too low to provide sufficient thermal coupling between the SOT and the sample surface. Above 1.2 K, the thermal sensitivity is ~1 μK/Hz$^{1/2}$ using either In, Pb, or Mo$_{66}$Re$_{34}$ SOTs, with the latter operating in magnetic fields up to 5 T. In order to achieve thermal imaging at lower temperatures, our design allows in-situ filling of the cell with liquid He. The magnetic and thermal imaging performance of the microscope immersed in superfluid He will be described elsewhere. Below we present examples of magnetic imaging of the antiferromagnetic topological insulator MnBi$_2$Te$_4$ and the ferromagnetic Weyl metal SrRuO$_3$, and thermal imaging in monolayer graphene in He gas.

MnBi$_2$Te$_4$ is the first-discovered intrinsic magnetic topological insulator, exhibiting the quantum anomalous Hall effect and axion insulator states in atomically thin flakes[48–50]. It has a layered structure, exhibiting A-type antiferromagnetism with MnTe ferromagnetic sheets stacked antiferromagnetically along the *c*-axis and Bi$_2$Te$_3$ sheets acting as spacers. As a result, in flakes with thickness down to several layers, the total magnetization should depend on whether the number of layers (*L*) is even or odd, which can be verified by nanoscale magnetic imaging. Figure 4a shows an AFM image of a thin flake of MnBi$_2$Te$_4$ with several terraces. The number of the septuple layers in the different regions derived from the measured thickness is indicated. An image of the local *dc* magnetic field $B_{dc}(x,y)$ at 4.6 K in the presence of $B_a = 0.7$ T is shown in Fig. 4b. It is clearly seen that in the odd-layer regions (*L* = 13, 23, 21 and 17), a strong magnetic signal is present (dark), while in the even-layer regions (*L* = 18 and 20), no net magnetization is visible, except for some small and isolated magnetic islands. The fine magnetic structure can be resolved more clearly by inspecting the *ac* magnetic signal measured at the TF resonance frequency, $B_{ac}^{TF}(x,y) = y_{ac}\partial B_{dc}(x,y)/\partial y$, where $y_{ac}$ is the amplitude of the SOT oscillation along the $y$ axis induced by the TF. This signal shows the spatial gradient of $B_{dc}$ with a higher signal-to-noise ratio due to the lower $1/f$ noise at high frequency. Our data presents a direct visualization of the even-odd effect and shows that the A-type antiferromagnetism persists in thin flakes, consistent with previous reports[51]. A quantitative analysis of the data will be presented elsewhere.

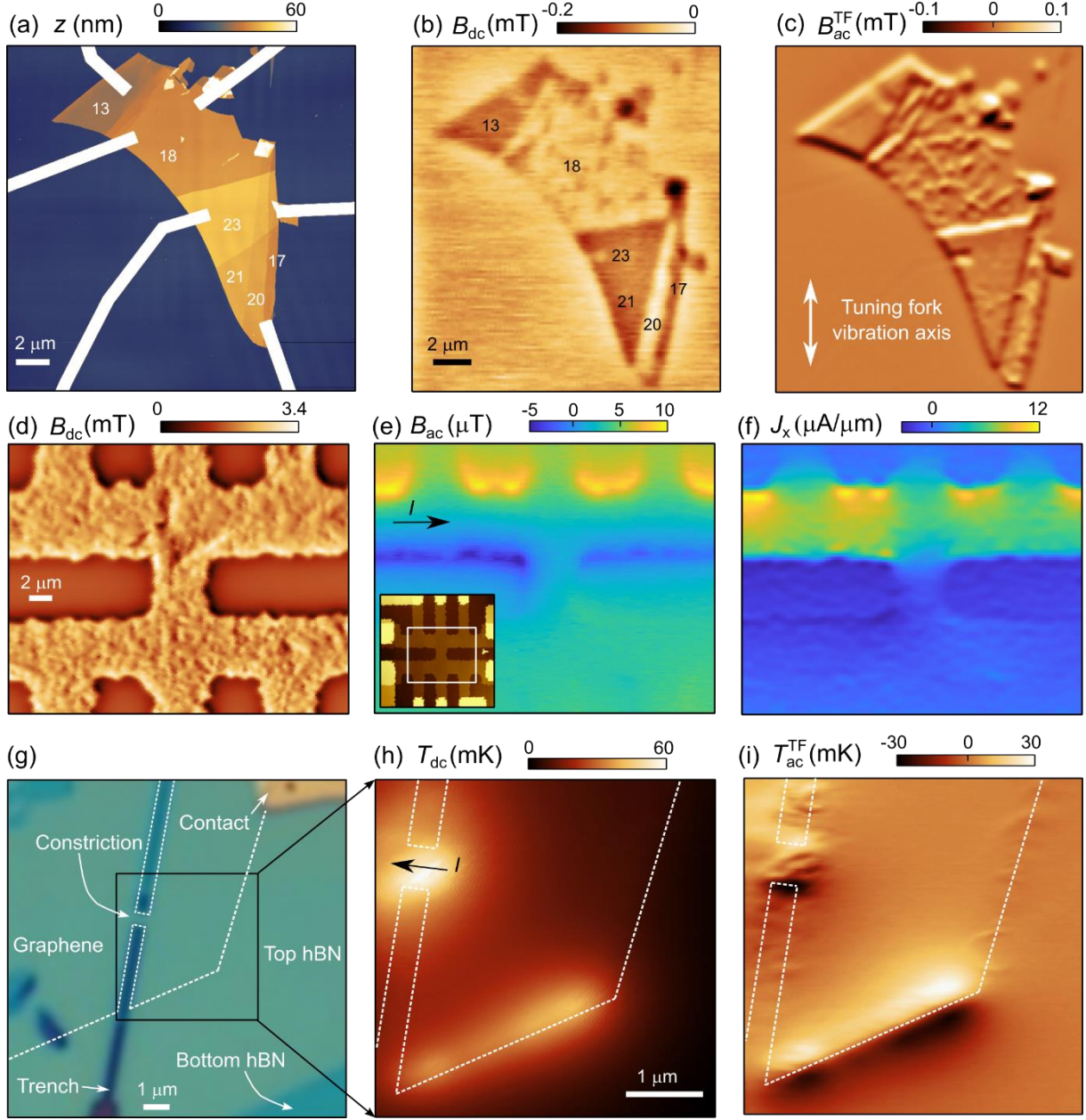

**Fig. 4**. (a) Topography of the MnBi$_2$Te$_4$ device measured using an atomic force microscope, with indicated number of septuple layers in the different regions. The white strips are Au contacts to the sample. (b) Image of the local magnetic field $B_{dc}(x,y)$ acquired with a Pb SOT at 4.6 K in applied magnetic field $B_a = 0.7$ T showing strong out-of-plane magnetism in the odd-layer regions and absence of net magnetization in the even-layer domains. (c) The local *ac* signal $B_{ac}^{TF}(x,y)$ at the TF resonance frequency acquired simultaneously with the *dc* signal. The $B_{ac}^{TF}(x,y)$ is proportional to the spatial derivative of the $B_{dc}(x,y)$ along the direction of the TF oscillation. (d) $B_{dc}(x,y)$ magnetic

field image of the ferromagnetism in SrRuO$_3$ film. (e) The Biot-Savart field $B_{ac}(x,y)$ induced by the *ac* current applied to the sample. Inset: AFM topographic image of the patterned film and the contacts. The white rectangle marks the imaged area. (f) Reconstructed distribution of the 2D current density component $J_x$ along the horizontal axis. (g) Optical micrograph of a graphene device with an etched constriction. The edges of the graphene are marked by the white dashed lines. (h) Thermal image $T_{dc}(x,y)$ showing local and nonlocal dissipation due to a *dc* current $I_{dc} = 3$ µA flowing through the constriction when the sample is biased at the charge neutrality point. Strong nonlocal dissipation is present along the bottom edge of graphene. (i) The *ac* signal $T_{ac}^{TF}(x,y)$ at the TF resonance frequency acquired simultaneously with the *dc* signal, revealing weak ring-like features due to resonant inelastic electron scattering off atomic defects along the right edge while strong featureless dissipation is observed along the bottom edge.

The itinerant ferromagnet SrRuO$_3$ has attracted broad interest due to the emergence of the topological Hall effect[52] and a topological band structure[53,54]. However, their relation to the widely discussed skyrmions is still under intense debate[55–59]. The SrRuO$_3$ also exhibits the sign-flip anomalous Hall effect in different crystalline phases[58]. Real-space observation of the topological magnetic texture and current-induced spin dynamics can provide valuable insight into the underlying mechanisms. Figure 4d shows the $B_{dc}(x,y)$ magnetic image of a patterned 12 nm thick SrRuO$_3$ film on SrTiO$_3$ (001) substrate in an applied magnetic field of 20 mT at 4.6 K. Strong magnetization with some inhomogeneity is clearly observed, evidencing ferromagnetism. Figure 4e shows the simultaneously imaged Biot-Savart field $B_{ac}(x,y)$ induced by an *ac* current of 20 µA applied along the top bar. The current distribution, reconstructed from $B_{ac}$ following the inversion procedure described in Meltzer *et al.*[60], is presented in Fig. 4f.

Figure 4g shows an optical image of a graphene sample patterned into a narrow constriction. Driving a *dc* current $I_{dc} = 3$ µA through the sample gives rise to enhanced dissipation in the constriction as shown by the thermal image $T_{dc}(x,y)$ in Fig. 4h, which was acquired using a Mo$_{66}$Re$_{34}$ Ø60 nm SOT in zero applied field at 4.6 K with He pressure in the cell of about 1 mbar. In addition to the local dissipation in the constriction, very strong nonlocal dissipation is observed along the bottom graphene edge. Interestingly, no dissipation is visible along the right edge. The fine structure of the dissipation can be resolved by inspecting the thermal signal at the TF frequency, $T_{ac}^{TF}(x,y) = y_{ac} \partial T_{dc}(x,y)/\partial y$, shown in Fig. 4i. Weak ring-like features due to resonant inelastic electron scattering off atomic defects at graphene edges[14,19] are visible along the right and left edges, while a strong broad dissipation is present along the bottom edge. One possibility is that at the bottom edge, which was cut using an AFM tip, the graphene might be folded, forming a narrow strip of bilayer graphene, which has a stronger electron-phonon coupling compared to monolayer graphene[61] and hence stronger dissipation.

**Conclusions**

We have demonstrated the operation of a scanning SQUID-on-tip microscope in a top-loading cryogen-free DR with an acceptable level of vibrations. The unique design, in which the microscope is encapsulated in a vacuum-tight cell with in-situ variable He gas pressure, allows variable-temperature thermal and magnetic imaging in gas, normal liquid and superfluid He environments. The introduced method of determining the relative vibrations between the tip and the sample based on local thermal gradients allows to analyze and quantify the vibration characteristics for further improvements of the system stability. The developed scanning SQUID-on-tip microscope opens a route for nanoscale spatially-resolved investigation of dissipation mechanisms, magnetism, and thermal and electronic transport in quantum materials at millikelvin temperatures and in the presence of vector magnetic fields.

**ACKNOWLEDGMENTS**

This work was supported by the European Research Council (ERC) under the European Union's Horizon research and innovation program (grant No 101089714), by the Israel Science Foundation grant No 687/22, and by the National Research Foundation (NRF) of Singapore and Israel Science Foundation (ISF) under ISF-NRF joint program (grants No 3518/20 and NRF2020-NRF-ISF004-3518). E.Z. acknowledges the support of the Andre Deloro Prize for Scientific Research and Leona M. and Harry B. Helmsley Charitable Trust grant #2112-04911.

**AUTHOR DECLARATIONS**

**Conflict of Interest**

The authors have no conflicts to disclose.

**DATA AVAILABILITY**

The data that support the findings of this study are available from the corresponding author upon request.